\documentclass[11pt]{article}

\usepackage[T1]{fontenc}
\usepackage[cp1251]{inputenc}
\usepackage{textcomp}
\usepackage[centertags]{amsmath}
\usepackage[mediummath]{nccmath}
\usepackage{amsfonts}
\usepackage{amssymb}
\usepackage[pdftex]{hyperref}
\usepackage{graphicx}
\usepackage{graphbox}
\usepackage[numbers,sort&compress]{natbib}

\usepackage{paperinitial}

\paperinitialization{15mm}{15mm}{15mm}{15mm}{2pt}{10pt}

\DeclareMathOperator{\re}{Re}
\DeclareMathOperator{\im}{Im}

\DeclareMathOperator{\arsh}{arsh}

\DeclareMathOperator{\Sp}{Sp}

\newcommand{\lan}{\langle}
\newcommand{\ran}{\rangle}
\newcommand{\bs}{\boldsymbol}

\newcommand{\e}{\varepsilon}
\newcommand{\vf}{\varphi}
\newcommand{\vk}{\varkappa}
\newcommand{\s}{\sigma}

\newcommand{\al}{\alpha}
\newcommand{\be}{\beta}
\newcommand{\ga}{\gamma}

\newcommand{\de}{\delta}
\newcommand{\De}{\Delta}

\newcommand{\la}{\lambda}

\newcommand{\spx}{\mathbf{x}}
\newcommand{\spy}{\mathbf{y}}

\newcommand{\spp}{\mathbf{p}}
\newcommand{\spk}{\mathbf{k}}
\newcommand{\spK}{\mathbf{K}}
\newcommand{\spe}{\mathbf{e}}
\newcommand{\spn}{\mathbf{n}}

\begin{document}
\allowdisplaybreaks[4]
\frenchspacing

\title{{\Large\textbf{Susceptibility of a single photon wave packet}}}

\date{}

\author{P.O. Kazinski\thanks{E-mail: \texttt{kpo@phys.tsu.ru}}\;\, and T.V. Solovyev\thanks{E-mail: \texttt{nevybrown@mail.ru}}\\[0.5em]
{\normalsize Physics Faculty, Tomsk State University, Tomsk 634050, Russia}
}

\maketitle

\begin{abstract}

The explicit compact expression for the susceptibility tensor of a single photon wave packet on the photon mass-shell is derived. It is assumed that the probe photon is hard, the test photon is soft, and their total energy is below the electron-positron pair creation threshold. It turns out that a single photon wave packet can be regarded as a birefringent gyrotropic dispersive medium in the process of light-by-light scattering. The explicit expression for the inclusive probability to record the probe photon in the process of light-by-light scattering is obtained in the first nontrivial order of perturbation theory where the interference effect of the free passed and scattered parts of the photon wave function dominates. This effect is of order $\al^2$ in contrast to the standard contribution to the light-by-light scattering cross-section which is of order $\al^4$. The possible nontrivial shapes of the wave functions of probe and test photons are taken into account. The evolution of the Stokes parameters of a probe photon is described. The change of the Stokes parameters is rather large for hard probe photons and sufficiently intense beams of soft test photons.

\end{abstract}

\section{Introduction}

The study of the properties inherent to elementary particles such as mass, spin, charges, magnetic and dipole moments, and others is one of the fundamental problems of physics. It was shown in the paper \cite{KazSol22} that another one such characteristics of particles is their susceptibility. Staying in line with traditions of classical physics, it appears at first sight that the susceptibility is a property of a group of particles or of particles with nontrivial internal structure. Nevertheless, as was shown in \cite{KazSol22}, the susceptibility tensor can be defined, evaluated, and measured experimentally for the wave packet of a single electron. In the present paper, we continue the investigation of susceptibilities of elementary particles and find the susceptibility tensor for the wave packet of a single photon on the photon mass-shell.

The simplest way to calculate the susceptibility tensor of a single photon wave packet could be in the use of the Heisenberg-Euler Lagrangian \cite{BBBB70,MarShuk06,FMST07,BatRiz13,KarbShai15,KarbMos18,Karb18,KarbMos20,Karb20,GiesKarbKlar22,KUMZ22,Ahmadin22,FIKKSTT}. This is the most common method to describe the light-by-light scattering process that allows one to obtain the effective susceptibility of a beam of photons or of a macroscopic electromagnetic field. However, in applying this procedure to derivation of the susceptibility tensor of a single photon, it is not immediately clear what should be taken as a background field since the average values of the electromagnetic field operator over Fock states vanish. Another drawback of this approach is that it implies the total energy of the probe and tested photons is much less than the electron-positron pair creation threshold and it is not applicable near this threshold. It is known (see, e.g., \cite{LandLifQED}) that the light-by-light cross section strongly depends on the energies of scattered particles and it rapidly increases near the electron-positron pair creation threshold. The second nonperturbative method to find the susceptibility tensor is to employ the exact expression for the photon polarization tensor on a strong plane wave electromagnetic background \cite{BeckMit75,BaiMilStr75,DHIMT14,DHIMT141,BMKP17}. This approach is restricted to plane electromagnetic tested waves and, as in the case of the Heisenberg-Euler effective action, is not immediately applicable to a wave packet of a single tested photon. Therefore, in the present paper we stick to the standard perturbative approach for description of light-by-light scattering \cite{KarpNeu50,KarpNeu51,Tollis64,Tollis65,ConTollPist71} fully taking into account the shapes of the wave functions of probe and tested photons.

By now there are papers where the influence of profiles of the wave packets of scattered photons on various aspects of the light-by-light scattering was studied \cite{Varf66,BMKP17,DHIMT14,DHIMT141,KarbShai15,KarbMos18,Karb18,KarbMos20,Karb20,GiesKarbKlar22,KUMZ22,FIKKSTT}. Nevertheless, as far as we known, the expression for the susceptibility tensor of a single photon wave packet and of a beam of photons of a general profile has not been found in a closed and concise form. In the present paper we fill this gap. Furthermore, we derive the explicit expression for the inclusive probability to record a probe photon in the light-by-light scattering taking into account the interference of the free passed part of the probe photon wave function with its scattered part. This interference effect stems from a change of the probe photon wave function in scattering on an effective medium with susceptibility tensor of the tested photon or of a beam of such photons. The inclusive probability depends on the nontrivial structure of the states of probe and tested photons. Under certain approximations, the expression for this probability implies a simple equation for evolution of the Stokes parameters of the probe photon that generalizes the relations obtained in \cite{KotSerb97,Maishev97,Sawyer04,BelMai12,Sawyer15}. It turns out that the evolution of the Stokes parameters of the probe photon depends severely on the shape of its wave packet, in particular, on the presence of imaginary part of the density matrix of its state in the momentum space. Unpolarized states of probe photons possessing the density matrix with nonzero imaginary part become polarized as a result of scattering on tested photons. The magnitude of the interference effect and, respectively, a change of the Stokes parameters can be rather large for scattering of the hard probe photon by a lengthy intense laser beam. This effect can be observed on the existing and planned experimental facilities provided the profile of the wave packet of a probe photon and its polarization can be controlled \cite{NakHom17,Takahashi18,EMRY22,Karb22,Budker22}.

The paper is organized as follows. In Sec. \ref{GenForm}, the general formula for the inclusive probability to record a probe photon scattered by a tested photon is given. Section \ref{Suscept_Phot} is devoted to derivation of the concise explicit expression for the susceptibility tensor of a single photon wave packet. Here we also provide the estimates for the order of magnitude of this quantity in different regimes. In the next Sec. \ref{Inclus_Probab}, we simplify the general expression for the inclusive probability and describe the evolution of the Stokes parameters of the probe photon. In Conclusion, we summarize the results. Some calculations arising in evaluating the inclusive probability are removed to Appendix \ref{App_Traces}. In Appendix \ref{Sus_Sing_Electr}, we generalize the expression for the susceptibility tensor of a single electron wave packet obtained in \cite{KazSol22} to a nonstationary case.

We follow the notation adopted in \cite{KazSol22}. The Greek indices $\al$, $\be$, $\bar{\al}$, $\bar{\be}$, $\ldots$ denote the quantum numbers of particle states. The Greek $\mu$ is the space-time index taking the values $\overline{0,3}$ and the Latin $i$, $j$ are the spatial indices. The Greek $\la=\pm1$ specifies the circular polarization, whereas $l,l'=\{1,2\}$ are for the linear polarization. The summation (integration) over repeated indices is always understood unless otherwise stated. We also suppose that the quantum states of particles are normalized to unity in some sufficiently large volume $V$. The complex conjugation is denoted by the bar over the symbol. Furthermore, wherever it does not lead to misunderstanding, we use the matrix notation. For example,
\begin{equation}
    \bar{a}a\equiv \bar{a}_\al a_\al\equiv\sum_\al \bar{a}_\al a_\al,\qquad \bar{d}Dd\equiv \bar{d}_{\bar{\al}} D_{\bar{\al}\al} d_{\al}= \sum_{\bar{\al},\al} \bar{d}_{\bar{\al}} D_{\bar{\al}\al} d_{\al},\qquad \text{etc.}
\end{equation}
The operators acting in the Fock space are denoted by letters with carets. We use the system of units such that $\hbar=c=1$ and $e^2=4\pi\al$, where $\al$ is the fine structure constant. The Minkowski metric is taken with the mostly minus signature.

\section{General formulas}\label{GenForm}

Consider the process of an elastic scattering of a photon by a photon in the leading nontrivial order of perturbation theory. As the initial state of photons at $t=t_1\rightarrow-\infty$, we take the coherent state defined by the density matrix
\begin{equation}\label{R_ph}
    \hat{R}_{ph}=|d\ran\lan \bar{d}| e^{-\bar{d}d},
\end{equation}
where $d_\al$ is the complex amplitude of the coherent state at the instant of time $t_1$. We suppose that the quantum numbers $\al$ contain the particle energy and
\begin{equation}
    d_\al= s_\al+h_\al,
\end{equation}
where $s_\al$ describes the state of the laser beam comprised of low energy photons and $h_\al$ determines the state of hard probe photons. Furthermore, we assume that the total energy of any two photons from the state $\hat{R}_{ph}$ is not enough to create an electron-positron pair, i.e., $s<4m^2$. The initial state of the whole system takes the form
\begin{equation}\label{dens_matr_ini}
    \hat{R}=\hat{R}_{ph}\otimes |0\ran_{e^-}\lan0|_{e^-}\otimes|0\ran_{e^+}\lan0|_{e^+},
\end{equation}
where $|0\ran_{e^-}$ is the vacuum state of electrons and $|0\ran_{e^+}$ is the vacuum state of positrons.

In order to define the quantum measurement in the final state at $t=t_2\rightarrow+\infty$, we introduce the projectors
\begin{equation}\label{projector}
    \hat{\tilde{\Pi}}_D:=1-\hat{\Pi}_D,\qquad \hat{\Pi}_D:=:\exp(-\hat{c}^\dag D\hat{c}):,
\end{equation}
where $D=D^\dag$ is the projector in the one-particle Hilbert space and $\hat{c}^\dag_\al$ and $\hat{c}_\al$ are the creation and annihilation operators for photons. The projector $\hat{\tilde{\Pi}}_D$ singles out the states in the Fock space that contain at least one photon with quantum numbers specified by the projector $D$ due to the fact that
\begin{equation}
    (D\hat{c})_\al \hat{\Pi}_D= \hat{\Pi}_D (\hat{c}^\dag D)_\al=0.
\end{equation}
Then the inclusive probability to record a photon by the detector at the instant of time $t_2$ reads
\begin{equation}\label{probab_D}
    P_D=\Sp(\hat{R} \hat{U}_{t_1,t_2}\hat{\tilde{\Pi}}_D\hat{U}_{t_2,t_1})=\Sp(\hat{R}(t_1)\hat{S}_{t_1,t_2}\hat{\tilde{\Pi}}_{D(t_2)} \hat{S}_{t_2,t_1}),
\end{equation}
where
\begin{equation}
    D_{\bar{\al}\al}(t_2)=D_{\bar{\al}\al}e^{i(k_{0\bar{\al}}-k_{0\al})t_2},
\end{equation}
and $\hat{R}(t_1)$ has the form \eqref{R_ph}, \eqref{dens_matr_ini}, where one should substitute
\begin{equation}\label{d(t)}
    d_\al\rightarrow d_\al(t)\Big|_{t=0}=d_\al e^{it_1k_{0\al}}.
\end{equation}
In expression \eqref{probab_D}, we have also introduced the standard notation for the evolution operator $\hat{U}_{t_2,t_1}$ and the $S$-operator.

Further we assume that $D_{\bar{\al}\al}$ is diagonal with respect to the photon energy and, consequently, $D_{\bar{\al}\al}(t_2)=D_{\bar{\al}\al}$. Moreover, it is convenient to specify the form of the complex amplitude $d_\al$ at $t=0$
and not at the initial instant of time $t_1$. Then $d_\al$ taken at the initial instant of time is found by reversing formula \eqref{d(t)}. Henceforth, for brevity, we denote the complex amplitude of the coherent state at the instant of time $t=0$ as $d_\al$. Bearing this in mind and taking the limits $t_2\rightarrow+\infty$, $t_1\rightarrow-\infty$, we obtain
\begin{equation}\label{probab_D1}
    P_D=\Sp(\hat{R}\hat{S}^\dag\hat{\tilde{\Pi}}_{D} \hat{S}),
\end{equation}
where $\hat{S}$ is the operator of the $S$-matrix.

When the aforementioned restrictions on the energies of photons in the initial state are satisfied, only the process of light-by-light scattering may happen in the leading order of perturbation theory. Then the $S$-matrix becomes
\begin{equation}
    \hat{S}=1+\hat{C}+\cdots,
\end{equation}
where the operator
\begin{equation}
    \hat{C}=\hat{c}^\dag_{\bar{\al}}\hat{c}^\dag_{\bar{\be}} C_{\bar{\al}\bar{\be}\al\be} \hat{c}_{\al}\hat{c}_{\be}
\end{equation}
describes the light-by-light scattering in the leading order of perturbation theory. In virtue of unitarity of the $S$-matrix,
\begin{equation}
    \hat{C}^\dag=-\hat{C},
\end{equation}
in the given order of perturbation theory and domain of parameters. Therefore,
\begin{equation}\label{C_sym}
    C_{\bar{\al}\bar{\be}\al\be}=C_{\bar{\be}\bar{\al}\al\be}=C_{\bar{\al}\bar{\be}\be\al}=-C^*_{\al\be\bar{\al}\bar{\be}}.
\end{equation}
At the same order of perturbation theory,
\begin{equation}\label{PD}
    P_D=\Sp(\hat{R}_{ph}\hat{\tilde{\Pi}}_{D})+ \big[\Sp(\hat{R}_{ph}\hat{\tilde{\Pi}}_{D}\hat{C})+c.c.\big]+\cdots.
\end{equation}
The traces of operators appearing in this expression are readily evaluated (see Appendix \ref{App_Traces})
\begin{equation}\label{traces}
\begin{split}
    \Sp(\hat{R}_{ph}\hat{\tilde{\Pi}}_{D})&=1-e^{-\bar{d}Dd},\\
    \Sp(\hat{R}_{ph}\hat{\tilde{\Pi}}_{D}\hat{C})&=\big[\bar{d}_{\bar{\al}} \bar{d}_{\bar{\be}} -(\bar{d}\tilde{D})_{\tilde{\al}}(\bar{d}\tilde{D})_{\bar{\be}} e^{-\bar{d}Dd} \big]C_{\bar{\al}\bar{\be}\al\be}d_\al d_\be,
\end{split}
\end{equation}
where $\tilde{D}_{\bar{\al}\al}:=\de_{\bar{\al}\al}-D_{\bar{\al}\al}$.

We assume that
\begin{equation}
    (Ds)_\al=0,
\end{equation}
i.e., the detector does not record soft photons $s_\al$. In this case,
\begin{equation}
    (\bar{d}\tilde{D})_\al=\bar{s}_\al+(\bar{h}\tilde{D})_\al,\qquad \bar{d}Dd=\bar{h}Dh.
\end{equation}
Furthermore, we suppose that the state of hard photons is close to a one particle Fock state and so we seek for a leading nontrivial contribution to \eqref{PD} in limit $h_\al\rightarrow0$. Then
\begin{equation}\label{PD1}
\begin{split}
    P_D=\,&\bar{h}Dh+\Big\{\big[(\bar{h}Dh)\bar{s}_{\bar{\al}}\bar{s}_{\bar{\be}}s_\al s_\be\\
    &+\big(\bar{h}_{\bar{\al}}\bar{h}_{\bar{\be}} -(\bar{h}\tilde{D})_{\bar{\al}} (\bar{h}\tilde{D})_{\bar{\be}} +2\bar{s}_{\bar{\al}}(\bar{h}D)_{\bar{\be}}\big)s_\al s_\be +4\bar{s}_{\bar{\al}}(\bar{h}D)_{\bar{\be}}s_\al h_\be\big]C_{\bar{\al}\bar{\be}\al\be}+c.c.\Big\}+\cdots.
\end{split}
\end{equation}
It follows from the property \eqref{C_sym} that the first term in the square brackets is purely imaginary. Hence its contribution is equal to zero. The next terms embraced by the parentheses vanish due to the energy conservation law and the assumption that the energies of photons in the state $s_\al$ are small in comparison with the energies of photons in the state $h_\al$. Then within the order of perturbation theory we consider, we can write
\begin{equation}\label{PD_fin}
    P_D=\bar{h}_{t}Dh_{t},\qquad (h_t)_{\bar{\be}}=h_{\bar{\be}}+\Phi_{\bar{\be}\be}h_\be,
\end{equation}
where
\begin{equation}
    \Phi_{\bar{\be}\be}:=4\bar{s}_{\bar{\al}}C_{\bar{\al}\bar{\be}\al\be}s_\al,
\end{equation}
and for conciseness we have added the term to \eqref{PD_fin} of order $\alpha^4$. This term does not coincide with the standard contribution proportional to $\alpha^4$ neglected in \eqref{PD}.

Formula \eqref{PD_fin} says that the detector records photons in the state $(h_t)_{\bar{\be}}$ that results from scattering of photons in the state $h_\be$ by the photons in the laser beam described by the state $s_\al$. In the case of a mixed initial state of probe photons with the density matrix $\rho_{\be\be'}$, the inclusive probability \eqref{PD_fin} is written as
\begin{equation}\label{PD_fin_mix}
    P_D=D_{\bar{\be}'\bar{\be}} (\de_{\bar{\be}\be}+\Phi_{\bar{\be}\be}) (\de_{\bar{\be}'\be'}+\bar{\Phi}_{\bar{\be}'\be'}) \rho_{\be\be'}.
\end{equation}
Let us stress that formulas \eqref{PD}, \eqref{PD1}, \eqref{PD_fin}, and \eqref{PD_fin_mix} do not contain the standard contribution defining the differential cross-section of light-by-light scattering because it is of a higher order with respect to the coupling constant. The standard contribution becomes the leading one in the domain of quantum numbers $\bar{\be}$ where $(Dh)_{\bar{\be}}$ is negligible. However, when the free passed probe photon wave function overlaps with its scattered part the interference contribution taken into account in \eqref{PD}, \eqref{PD1}, \eqref{PD_fin}, and \eqref{PD_fin_mix} is stronger by four orders of magnitude than the standard one.

\section{Susceptibility of a photon}\label{Suscept_Phot}

The above general formulas allow one to deduce the susceptibility of a single photon wave packet on the mass-shell and to find the inclusive probability to record a photon scattered by other photon or by the laser beam of photons. The amplitude of light-by-light scattering is given in \cite{KarpNeu50,KarpNeu51,Tollis64,Tollis65,ConTollPist71,LandLifQED}. In our notation,
\begin{equation}
\begin{gathered}
    \al=(\la_1,\spk_1),\qquad\be=(\la_2,\spk_2),\qquad\bar{\al}=(\la_3,\spk_3),\qquad \bar{\be}=(\la_4,\spk_4),\\
    \sum_\be=\sum_{\la_2}\int\frac{Vd\spk_2}{(2\pi)^3},\qquad h_\be=\sqrt{\frac{(2\pi)^3}{V}} h_{\la_2}(\spk_2).
\end{gathered}
\end{equation}
The normalization condition takes the form
\begin{equation}
    \sum_\al\bar{h}_\al h_\al=\sum_{\la}\int d\spk |h_{\la}(\spk)|^2=1,\qquad \sum_{\la}\int d\spk |s_{\la}(\spk)|^2=N_s,
\end{equation}
where $N_s$ is the average number of photons in the beam $s_\al$. The circular polarization vectors are defined as \cite{KarpNeu50,KarpNeu51,Tollis64,Tollis65,ConTollPist71}
\begin{equation}
    \spe^{(\la)}(\spk)=\frac{1}{\sqrt{2}}(\spe_1(\spk)+i\la\spe_2(\spk)),
\end{equation}
where $\la=\pm1$, the linear polarization vector $\spe_1(\spk)$ is perpendicular to the reaction plane, the linear polarization vector $\spe_2(\spk)$ lies in the reaction plane, and $\{\spe_1(\spk),\spe_2(\spk),\spk\}$ constitute a right-handed triple.

In this case,
\begin{equation}\label{phi_bb}
    \Phi_{\bar{\be}\be}=\frac{\pi i}{2V}\sum_{\la_1,\la_3}\int d\spk_1 d\spk_3\de(k_3+k_4-k_1-k_2) \bar{s}_{\la_3}(\spk_3) s_{\la_1}(\spk_1) \frac{M_{\la_3\la_4\la_1\la_2}(s,t,u)}{\sqrt{k_0(\spk_1)k_0(\spk_2)k_0(\spk_3)k_0(\spk_4)}},
\end{equation}
where
\begin{equation}
\begin{gathered}
    s=(k_1+k_2)^2=(k_3+k_4)^2=2k_1k_2=2k_3k_4,\qquad t=(k_1-k_3)^2=(k_2-k_4)^2=-2k_1k_3=-2k_2k_4,\\
    u=(k_1-k_4)^2=(k_2-k_3)^2=-2k_1k_4=-2k_2k_3.
\end{gathered}
\end{equation}
In particular,
\begin{equation}
    s=k_3^0k_4^0(\spn_3-\spn_4)^2,
\end{equation}
where $\spn_{3,4}:=\spk_{3,4}/|\spk_{3,4}|$. It is clear that $s+t+u=0$. Integrating over the spatial momenta $\spk_1$ in \eqref{phi_bb}, we arrive at
\begin{equation}\label{Phi_k}
    \Phi_{\bar{\be}\be}=\frac{\pi i}{2V}\sum_{\la_1,\la_3}\int d\spk_3\de(k^0_3+k^0_4-k^0_1-k^0_2) \bar{s}_{\la_3}(\spk_3) s_{\la_1}(\spk_1) \frac{M_{\la_3\la_4\la_1\la_2}(s,t,u)}{\sqrt{|\spk_1||\spk_2||\spk_3||\spk_4|}}\Big|_{\spk_1=\spk_3+\spk_4-\spk_2}.
\end{equation}
Introduce the notation,
\begin{equation}
    s_{\la}(\spk;x^0):=e^{-ik_0(\spk)x^0}s_{\la}(\spk),
\end{equation}
and write the delta function expressing the energy conservation law as a Fourier transform. Then we have
\begin{equation}\label{Phi_t}
    \Phi_{\bar{\be}\be}=i \sum_{\la_1,\la_3}\int \frac{d\spk_3 dx^0}{4V}e^{i(k^0_4-k^0_2)x^0} \bar{s}_{\la_3}(\spk_3;x^0) s_{\la_1}(\spk_1;x^0) \frac{M_{\la_3\la_4\la_1\la_2}(s,t,u)}{\sqrt{|\spk_1||\spk_2||\spk_3||\spk_4|}}\Big|_{\spk_1=\spk_3+\spk_4-\spk_2}.
\end{equation}

In order to find the susceptibility tensor of a single photon wave packet on the mass-shell, we compare the scattering amplitude of the hard probe photon $(h_t)_{\bar{\be}}$ with the amplitude of scattering by a medium with a certain susceptibility tensor $\chi_{ij}$ in the first Born approximation. Let the medium possess the susceptibility tensor
\begin{equation}
    \chi_{ij}\big(\frac{x+y}{2},x-y\big),
\end{equation}
where $x=(x^0,\spx)$. The dependence of $\chi_{ij}$ on $(x+y)/2$ is supposed to be slow. The second argument of $\chi_{ij}$ characterizes the frequency and spatial dispersions, and $\chi_{ij}$ is a rapidly varying function of this argument. Then, in the first Born approximation, the amplitude of scattering of a photon by the medium with such a susceptibility tensor becomes (see, e.g., \cite{LandLifshECM,KazKor22})
\begin{equation}\label{S_matr_chi}
    S_{\ga'\ga}=\de_{\ga'\ga}+i\frac{k^{1/2}_{0\ga'}k^{1/2}_{0\ga}}{2V}\int d^4x \bar{e}^{(\la')}_i(\spk')\chi_{ij}(x;K)e_j^{(\la)}(\spk) e^{i(k_{0\ga'}-k_{0\ga})x^0-i(\spk'-\spk)\spx},
\end{equation}
where $K_\mu:=(k'_\mu+k_\mu)/2$ and
\begin{equation}
    \chi_{ij}(x;K):=\int d^4z e^{iK_\mu z^\mu}\chi_{ij}(x,z).
\end{equation}
It is useful to write \eqref{S_matr_chi} as
\begin{equation}\label{S_matr_chi1}
    S_{\ga'\ga}=\de_{\ga'\ga}+i\frac{k^{1/2}_{0\ga'}k^{1/2}_{0\ga}}{2V}\int dx^0 \bar{e}^{(\la')}_i(\spk')\tilde{\chi}_{ij}(x^0,\De\spk;K)e_j^{(\la)}(\spk) e^{i(k_{0\ga'}-k_{0\ga})x^0},
\end{equation}
where $\De\spk:=\spk'-\spk$ and we have introduce the notation for the Fourier transform of the susceptibility tensor with respect to the slowly varying spatial argument. Comparing \eqref{Phi_t} with \eqref{S_matr_chi1}, we obtain
\begin{equation}\label{suscept_gen}
    \tilde{\chi}_{ij}(x^0,\De\spk;K)=\sum_{\la_1,\la_3}\int \frac{d\spk_3\bar{s}_{\la_3}(\spk_3;x^0)s_{\la_1}(\spk_3+\De\spk;x^0)}{2|\spk_4||\spk_2||\spk_3|^{1/2}|\spk_3+\De\spk|^{1/2}} M_{\la_3\la_4\la_1\la_2}e^{(\la_4)}_i(\spk_4)\bar{e}^{(\la_2)}_j(\spk_2),
\end{equation}
where
\begin{equation}
    \spk_2=\mathbf{K}-\frac{\De\spk }{2},\qquad \spk_4=\mathbf{K}+\frac{\De\spk }{2}.
\end{equation}
Formula \eqref{suscept_gen} gives the general expression for the on-shell susceptibility tensor of photons in the state $s_\al$.

Let us simplify expression \eqref{suscept_gen}. Recall that \cite{KarpNeu50,KarpNeu51,Tollis64,Tollis65,ConTollPist71,LandLifQED}
\begin{equation}
    M_{\la_1\la_2\la_3\la_4}=M_{-\la_1,-\la_2,-\la_3,-\la_4},\qquad M_{\la_1\la_2\la_3\la_4}=M_{\la_3\la_4\la_1\la_2},\qquad M_{\la_1\la_2\la_3\la_4}=M_{\la_2\la_1\la_4\la_3}.
\end{equation}
In the limit of a small momentum transfer, $|t|\ll 4 m^2$, $|t|\ll s$, the nonvanishing independent amplitudes are written as
\begin{equation}\label{M_t0}
    M_{++++}(s)=M_{+-+-}(-s)=8\al^2 f(s),\qquad M_{++--}(s)=-8\al^2 g(s),
\end{equation}
where
\begin{equation}
\begin{split}
    f(s)&=-\Big[1+\Big(2-\frac{4}{s'}\Big)B(s') +\Big(-4+\frac{4}{s'}\Big)B(-s') +\Big(\frac{4}{s'}-\frac{8}{s'^2}\Big)T(s') +\Big(2-\frac{4}{s'}-\frac{8}{s'^2}\Big)T(-s') \Big]_{s'\rightarrow s/m^2},\\
    g(s)&=-\Big[1 +\frac{4}{s'}B(s') -\frac{4}{s'}B(-s') +\frac{8}{s'^2} T(s') +\frac{8}{s'^2}T(-s') \Big]_{s'\rightarrow s/m^2},
\end{split}
\end{equation}
and
\begin{equation}
    B(s)=\sqrt{1-\frac{4}{s}}\arsh \frac{\sqrt{-s}}{2}-1=\sqrt{\frac{4}{s}-1}\arcsin \frac{\sqrt{s}}{2}-1,\qquad T(s)=\arsh^2\frac{\sqrt{-s}}{2}=-\arcsin^2\frac{\sqrt{s}}{2},
\end{equation}
where the principal branches of multivalued functions are taken and $s\rightarrow s+i0$. For $s$, $|t|$, $|u|$ much less than $4m^2$, the independent amplitudes become
\begin{equation}
\begin{gathered}
    M_{++++}=\frac{11\al^2}{45 m^4}s^2,\qquad M_{+-+-}=\frac{11\al^2}{45 m^4}u^2,\\
    M_{+--+}=\frac{11\al^2}{45 m^4}t^2,\qquad M_{++--}=-\frac{\al^2}{15m^4}(s^2+t^2+u^2),\qquad M_{+++-}=0.
\end{gathered}
\end{equation}
In the general case, the explicit expressions for $M_{\la_1\la_2\la_3\la_4}$ are presented in \cite{KarpNeu50,KarpNeu51,Tollis64,Tollis65,ConTollPist71,LandLifQED}.

\begin{figure}[t]
   \centering
   \includegraphics*[width=0.5\linewidth]{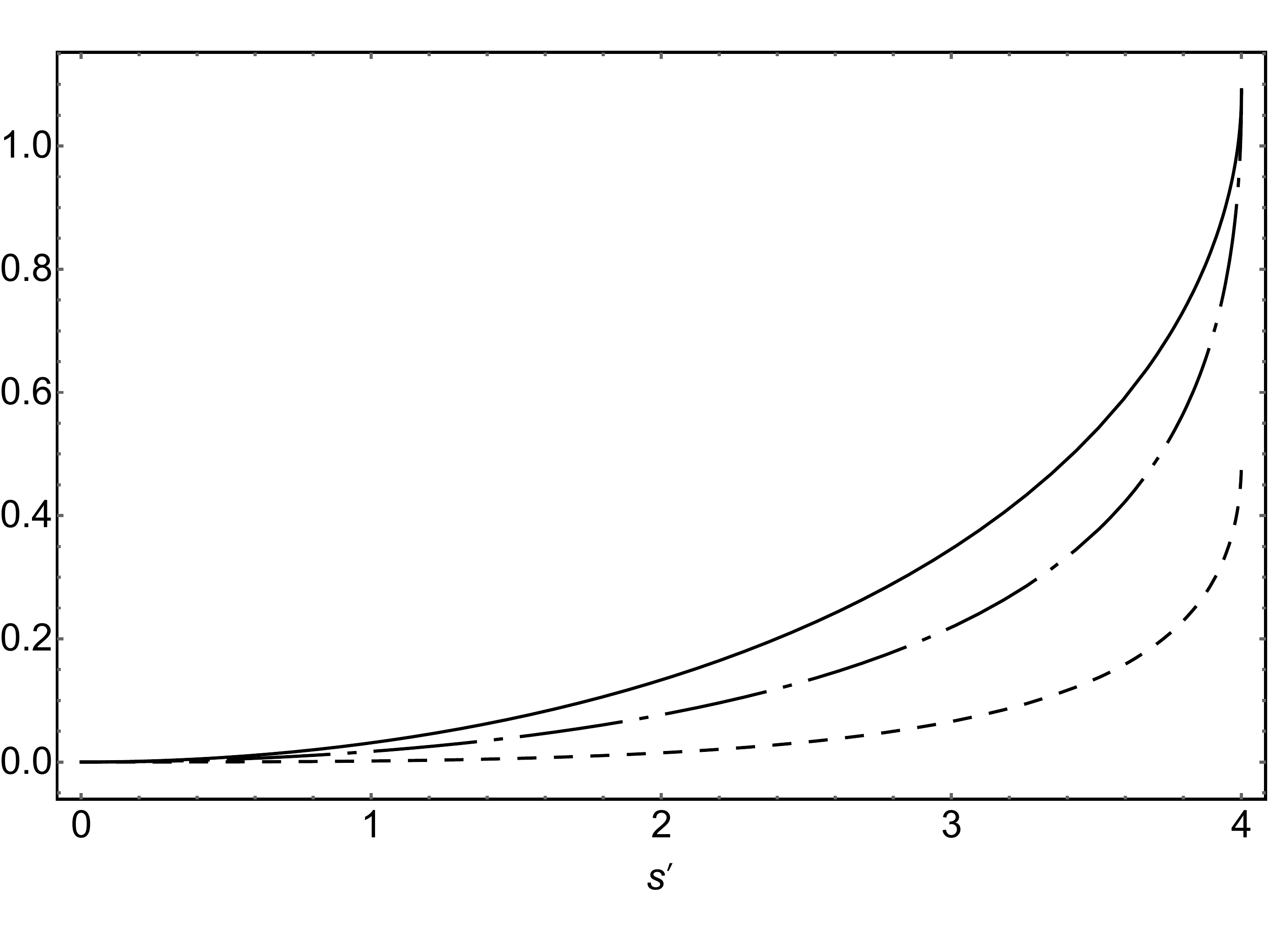}
   \caption{{\footnotesize The dependence of $f_s(s)$, $f_a(s)$, and $g(s)$ on $s'=s/m^2$. The solid line is $f_s(s)$, the dashed line is $f_a(s)$, and the dashed dotted line is $g(s)$.}}
\label{ffg_plots}
\end{figure}

In order to proceed, we assume that the states $s_\al$ and $h_\be$ are such that
\begin{equation}\label{condition1}
    |\De\spk|\ll|\spk_3|,\qquad |\De\spk|\ll|\spk_4|,
\end{equation}
where $\De\spk:=\spk_4-\spk_2$. In fact, condition \eqref{condition1} means that the dispersion of momenta in the wave packet of a hard probe photon is much less than the average energy of modes in the state $s_\al$. The second condition in \eqref{condition1} follows from the first one inasmuch as, by assumption, the energy of the photon $h_\be$ is much higher than the energies of the modes in the state $s_\al$. In this case, $s\approx |u|\ll 4m^2$ and $|t|\ll s$ and so one can use formulas \eqref{M_t0} for the invariant scattering amplitudes. In the leading order in $\De\spk$, we can discard $\De\spk$ in the integrand of \eqref{suscept_gen} while keeping the argument of the function $s_{\la_1}(\spk_3+\De\spk)$. In the general case, this function can vary rapidly even for a small deviation, $\De\spk$, of its argument. Then
\begin{equation}
    s=|\spk_3||\mathbf{K}|(\spn_3-\spn)^2,\qquad \spn:=\mathbf{K}/|\mathbf{K}|.
\end{equation}
Denoting concisely
\begin{equation}
    s_{\la_3\la_1}:=\bar{s}_{\la_3}(\spk_3;x^0)s_{\la_1}(\spk_3+\De\spk;x^0),
\end{equation}
we obtain
\begin{equation}
\begin{split}
    \sum_{\la_1,\la_3}s_{\la_3\la_1}M_{\la_3\la_4\la_1\la_2}e^{(\la_4)}_i(\mathbf{K})\bar{e}^{(\la_2)}_j(\mathbf{K}) =&\,8\al^2 \Big[f_s(s)(s_{++}+s_{--}) +f_a(s)(s_{++}-s_{--})\s_2\\
    &-g(s)\frac{s_{+-}+s_{-+}}{2}\s_3 +g(s)\frac{s_{+-}-s_{-+}}{2i}\s_1 \Big]_{ll'}(e_{l})_i(\spK) (e_{l'})_j(\spK)\\
    =&\,8\al^2 \Big[f_s(s)(s_{11}+s_{22}) +if_a(s)(s_{21}-s_{12})\s_2\\
    &-g(s)\frac{s_{11}-s_{22}}{2}\s_3 +g(s)\frac{s_{12}+s_{21}}{2}\s_1 \Big]_{ll'}(e_{l})_i(\spK) (e_{l'})_j(\spK)\\
    =&\,8\al^2 \Big[f_s(s) (s^\dag \tilde{s}) +f_a(s) (s^\dag\s_{2}\tilde{s})\s_2\\
    &-\frac{g(s)}2 \big((s^\dag\s_3 \tilde{s})\s_3 -(s^\dag\s_1 \tilde{s})\s_1\big) \Big]_{ll'}(e_{l})_i(\spK) (e_{l'})_j(\spK),
\end{split}
\end{equation}
where the basis of linear polarization vectors has been used, in the last equality we have rewritten the foregoing expression with the aid of sigma matrices, $\tilde{s}_l:=s_l(\spk_3+\De\spk;x^0)$, and
\begin{equation}
    f_s(s):=[f(s)+f(-s)]/2,\qquad f_a(s):=[f(s)-f(-s)]/2.
\end{equation}
Notice that $f_s(s)$, $f_a(s)$, and $g(s)$ are monotonically increasing functions for $s\in[0,4m^2]$ and are nonnegative on this interval. Moreover,
\begin{equation}\label{fsfag_asymp}
\begin{gathered}
    f_s(s)=\frac{11s^2}{360m^4}+\frac{13s^4}{21600m^8}+\cdots,\qquad f_a(s)=\frac{s^3}{630 m^6}+\frac{s^5}{17325m^{10}}+\cdots,\\
    g(s)=\frac{s^2}{60m^4}+\frac{s^4}{1890m^8}+\cdots,
\end{gathered}
\end{equation}
and
\begin{equation}
\begin{gathered}
    f_s(4m^2)=\frac{1}{2}\arsh^21+\frac{3\pi^2}{8}-3\approx1.0895,\qquad f_a(4m^2)=3\sqrt{2}\arsh1-\arsh^21-\frac{\pi^2}{4}\approx0.495,\\ g(4m^2)=\sqrt{2}\arsh1-\frac12 \arsh^21+\frac{\pi^2}{8}-1\approx1.0917.
\end{gathered}
\end{equation}
The plots of the functions $f_s(s)$, $f_a(s)$, and $g(s)$ are presented in Fig. \ref{ffg_plots}.

Introduce the relativistic coordinate representation of the complex amplitudes $s_\al$ as
\begin{equation}
    s_\la(x):=\int\frac{d\spk}{\sqrt{(2\pi)^32|\spk|}}e^{i\spk\spx}s_\la(\spk;x^0),
\end{equation}
and
\begin{equation}
\begin{split}
    \psi_{s,\la}(x)&:=f^{1/2}_s(s)s_\la(x)=\int\frac{d\spk_3 f^{1/2}_s(s)}{\sqrt{(2\pi)^32|\spk_3|}}e^{i\spk_3\spx}s_\la(\spk_3;x^0),\\
    \psi_{a,\la}(x)&:=f^{1/2}_a(s)s_\la(x)=\int\frac{d\spk_3 f^{1/2}_a(s)}{\sqrt{(2\pi)^32|\spk_3|}} e^{i\spk_3\spx}s_\la(\spk_3;x^0),\\
    \psi_{g,\la}(x)&:=g^{1/2}(s)s_\la(x)=\int\frac{d\spk_3 g^{1/2}(s)}{\sqrt{(2\pi)^32|\spk_3|}}e^{i\spk_3\spx}s_\la(\spk_3;x^0),
\end{split}
\end{equation}
where $f^{1/2}_s(s)$, $f^{1/2}_a(s)$, and $g^{1/2}_s(s)$ acting on $s_\la(x)$ are understood as pseudodifferential operators with $k^i_{3}=-i\partial/\partial x^i$. In that case, the susceptibility tensor can be cast into the form
\begin{equation}
    \chi_{ij}(x;K)=\frac{8\al^2}{\spK^2}\Big[(\psi_s^\dag\psi_s)\de^\perp_{ij} -i(\psi^\dag_a\s_2\psi_a)\e_{ijk}n_k -\frac{1}2 \big((\psi_g^\dag\s_3 \psi_g)\s_3 -(\psi_g^\dag\s_1 \psi_g)\s_1\big)_{ll'}(e_{l})_i(\spK) (e_{l'})_j(\spK)\Big].
\end{equation}
The last term in the square brackets can be simplified so that
\begin{equation}\label{ch_fin}
    \chi_{ij}(x;K)=\frac{8\al^2}{\spK^2}\Big\{\big[(\psi_s^\dag\psi_s)+\frac{1}{2}|\psi_{g,+}\psi_{g,-}|\big]\de^\perp_{ij} -i(\psi^\dag_a\s_2\psi_a)\e_{ijk}n_k -|\psi_{g,+}\psi_{g,-}| e^\vf_i(\spK) e^\vf_j(\spK)\Big\},
\end{equation}
where $\de_{ij}^\perp:=\de_{ij}-n_in_j$ and $e^\vf_i$ is the polarization vector $e_{1i}$ rotated by an angle of $\vf=-\arg(\psi_{g+}\psi_{g-})/2$ in the plane spanned by the vectors $\{\spe_1,\spe_2\}$. The susceptibility of a single photon wave packet is obtained when one retains the leading term in formulas \eqref{R_ph}, \eqref{PD_fin} for $s_\al\rightarrow0$. It is clear from these formulas that expression \eqref{ch_fin} also holds for the wave packet of a single photon, where $s_\la(\spk)$ should be interpreted as a single photon wave function.

The susceptibility tensor \eqref{ch_fin} corresponds to a birefringent gyrotropic dispersive medium. As is seen from asymptotics \eqref{fsfag_asymp}, the term related to gyrotropy is suppressed for $s\ll4m^2$, in particular, it is absent in the approach based on the Heisenberg-Euler Lagrangian. For infinitely small $|\spK|$, gyrotropy vanishes and the whole expression \eqref{ch_fin} tends to a finite nonzero limit. Furthermore, gyrotropy disappears in the case when $s_1=0$ or $s_2=0$. The last term in \eqref{ch_fin} vanishes for $s_+=0$ or $s_-=0$. In that case, the wave packet of a single photon is purely gyrotropic. Notice that for $s\approx 4m^2$ the contribution of the term responsible for gyrotropy of the wave packet is of the same order as the main contribution to the susceptibility tensor standing at $\de^\perp_{ij}$.

Let us estimate the magnitude of the susceptibility \eqref{ch_fin}. By the order of magnitude,
\begin{equation}
    \chi_{ij}\sim \frac{8\al^2}{\spK^2}f_s(s)\mathbf{A}^2 =\frac{2\al}{\pi}f_s(s) \frac{m^2}{\spK^2} K_u^2,
\end{equation}
where $\mathbf{A}(x)$ is the electromagnetic potential in the Coulomb gauge and
\begin{equation}
    K_u^2:= e^2\mathbf{A}^2/m^2
\end{equation}
is the undulator strength parameter characterizing the applicability of the standard perturbation theory \cite{BaKaStrbook,Bula96}. If $K_u\ll1$, then the perturbation theory is applicable, while for $K_u\gtrsim1$ the background field has to be taken into account nonperturbatively. Taking approximately, $s\sim 4|\spk_3||\spK|$, we have
\begin{equation}
    \chi_{ij}\sim \frac{32\al^2}{\pi} f_s(s) \frac{m^2\spk_3^2}{s^2} K_u^2.
\end{equation}
If the energy of the hard probe photon is close to the electron-positron pair creation threshold, $s\sim 4m^2$, then $f_s(s)\sim1$ and
\begin{equation}
    \chi_{ij}\sim \frac{2\al}{\pi} \frac{\spk_3^2}{m^2} K_u^2.
\end{equation}
For $s\ll4 m^2$, we have $f_s(s)\sim 11 s^2/360 m^4$ and
\begin{equation}\label{chi_est_low}
    \chi_{ij}\sim\frac{\al}{\pi} \frac{\spk_3^2}{m^2}K_u^2.
\end{equation}
Hence, we can use the same estimate for the susceptibility in the whole range of $s$ up to $s=4m^2$.

In order to find the magnitude of the susceptibility of a single photon wave packet, we can employ the estimate
\begin{equation}
    \mathbf{A}^2\sim n_s/|\spk_3|,
\end{equation}
where $n_s$ is the photon number density at a given point. Therefore,
\begin{equation}\label{chi_est_gen}
    \chi_{ij}\sim 8\al^2 f_s(s)\frac{n_s}{\spK^2|\spk_3|} =8\al^2 f_s(s)\frac{w_s}{\spK^2\spk_3^2},
\end{equation}
where $w_s$ is the energy density of photons in the state $s_\al$. By the order of magnitude, $n_s\sim\s_s^3$, where $\s_s$ is the standard deviation of momenta in the wave packet of a soft photon $s_\al$. Then
\begin{equation}
    \chi_{ij}\sim 128\al^2 \frac{f_s(s)}{s^2}|\spk_3|\s_s^3.
\end{equation}
As for the probe photon near the electron-positron pair creation threshold, $s\sim4m^2$, we obtain
\begin{equation}
    \chi_{ij}\sim 8\al^2 \frac{|\spk_3|\s_s^3}{m^4}\lesssim 8\al^2 \frac{\spk_3^4}{m^4},
\end{equation}
where we have taken $\s_s\approx |\spk_3|$ for the upper estimate. If $s\ll 4m^2$, then
\begin{equation}\label{chi_est}
    \chi_{ij}\sim4\al^2\frac{|\spk_3|\s_s^3}{m^4} \lesssim 4 \al^2 \frac{\spk_3^4}{m^4}.
\end{equation}
For example, the quantity on the right-hand side of \eqref{chi_est} is equal to $3.13\times 10^{-27}$ for the photon in the state $s_\al$ with the energy $1$ eV.

\section{Inclusive probability}\label{Inclus_Probab}

Let us find the explicit expression for the inclusive probability \eqref{PD_fin_mix}. From \eqref{Phi_k} we have
\begin{multline}\label{Phi_h}
    \Phi_{\bar{\be}\be}h_\be=i\sqrt{\frac{(2\pi)^3}{V}}\sum_{\la_1,\la_2,\la_3}\int \frac{d\spk_3 d\spk_2}{4(2\pi)^2}\de(k^0_3+k^0_4-k^0_1-k^0_2) \bar{s}_{\la_3}(\spk_3) s_{\la_1}(\spk_1)\times\\ \times\frac{M_{\la_3\la_4\la_1\la_2}(s,t,u) h_{\la_2}(\spk_2)}{\sqrt{|\spk_1||\spk_2||\spk_3||\spk_4|}}\Big|_{\spk_1=\spk_3+\De\spk}.
\end{multline}
To reveal the main features of expression \eqref{PD_fin_mix}, we assume that $|\De\spk|$ not only satisfies the conditions \eqref{condition1} but is much less than the typical scale of variation of the complex amplitude $s_\la(\spk)$. In the coordinate space, this condition means that the typical scale of variation of the wave function or of the average electromagnetic field of soft photons in the state $s_\al$ is much less than the diameter of the region of localization of the wave function of the hard probe photon $h_\be$. Notice that, in the plane-wave limit for the state $h_\be$, where $|\De\spk|\rightarrow0$, all the above conditions are satisfied.

Then one can neglect the dependence on $\De\spk$ and put $\spk_1\approx\spk_3$ and $\spk_2\approx\spk_4$ in all the functions appearing in the integrand of \eqref{Phi_h} apart from the delta function and $h_{\la_2}(\spk_2)=h_{\la_2}(\spk_4-\De\spk)$. As regards the argument of the delta function, we have
\begin{equation}
    k^0_3+k^0_4-k^0_1-k^0_2\approx(\spn_4-\spn_3)\De\spk.
\end{equation}
Introduce the splitting
\begin{equation}
    \spk_2=\spk_{2\parallel}+\spk_{2\perp},
\end{equation}
where
\begin{equation}
    \spk_{2\parallel}:=(\spn_4-\spn_3)\frac{(\spk_2 (\spn_4-\spn_3))}{(\spn_4-\spn_3)^2},
\end{equation}
and analogously for other vectors. Integrating the delta function in \eqref{Phi_h}, we come to
\begin{equation}\label{Phi_h1}
    \Phi_{\bar{\be}\be}h_\be=i\sqrt{\frac{(2\pi)^3}{V}}\sum_{\la_1,\la_2,\la_3}\int \frac{d\spk_3}{(2\pi)^2}  \frac{ \bar{s}_{\la_3}(\spk_3) s_{\la_1}(\spk_3)M_{\la_3\la_4\la_1\la_2} \tilde{h}_{\la_2}(\spk_{4\parallel})}{4|\spn_4-\spn_3||\spk_3||\spk_4|},
\end{equation}
where
\begin{equation}
    \tilde{h}_{\la_2}(\spk_{4\parallel}):=\int d\spk_{4\perp} h_{\la_2}(\spk_{4\parallel},\spk_{4\perp}).
\end{equation}

To simplify further the expression \eqref{Phi_h1}, we suppose that the complex amplitude $s_\al$ is such that the dispersion of the vector $\spn_3$ in this state is small, i.e., this state of photons is paraxial. Setting $\spn_3=\spn_{30}$, where $\spn_{30}$ is the average value of $\spn_3$ in the state $s_\al$, and using the approximate expressions for the invariant scattering amplitudes \eqref{M_t0}, the wave function of the hard probe photon after scattering \eqref{PD_fin} is given by
\begin{equation}
    (h_t)_{\bar{\be}}=\sqrt{\frac{(2\pi)^3}{V}}\big[h_{\la_4}(\spk_4) +i\vk \sum_{\la_2} (\xi_0+\bs\xi\bs\s)_{\la_4\la_2} \tilde{h}_{\la_2}(\spk_{4\parallel})\big],
\end{equation}
where
\begin{equation}\label{vk_xi}
\begin{split}
    \vk&=\frac{\al^2}{2\pi^2|\spk_4||\spn_4-\spn_{30}|},\\
    \xi_0&=\int \frac{d\spk_3}{|\spk_3|}f_s(s)  s^\dag(\spk_3)s(\spk_3),\\
    \xi_1&=-\int \frac{d\spk_3}{2|\spk_3|} g(s)  s^\dag(\spk_3)\s_1s(\spk_3),\\
    \xi_2&=\int \frac{d\spk_3}{2|\spk_3|} g(s)  s^\dag(\spk_3)\s_2s(\spk_3),\\
    \xi_3&=-\int \frac{d\spk_3}{|\spk_3|} f_a(s)  s^\dag(\spk_3)\s_3s(\spk_3),
\end{split}
\end{equation}
and $s=|\spk_3||\spk_4|(\spn_4-\spn_{30})^2$. Let us stress that expressions \eqref{vk_xi} are written in the chiral basis. As for the basis of linear polarization vectors $\spe_{1,2}$, the corresponding expressions take the form
\begin{equation}
\begin{split}
    \xi^l_0&=\int \frac{d\spk_3}{|\spk_3|}f_s(s)  s^\dag(\spk_3)s(\spk_3),\\
    \xi^l_1&=\int \frac{d\spk_3}{2|\spk_3|} g(s)  s^\dag(\spk_3)\s_1s(\spk_3),\\
    \xi^l_2&=\int \frac{d\spk_3}{|\spk_3|} f_a(s)  s^\dag(\spk_3)\s_2s(\spk_3),\\
    \xi^l_3&=-\int \frac{d\spk_3}{2|\spk_3|} g(s)  s^\dag(\spk_3)\s_3s(\spk_3),
\end{split}
\end{equation}
where $s_l(\spk_3)$ are also given in the basis of linear polarization vectors. In particular, if $s_1=0$ or $s_2=0$, then $\xi^l_2=0$. If $s_+=0$ or $s_-=0$, then $\xi^l_1=\xi^l_3=0$.

The formulas above are easily generalized to the case where the initial state of the probe photon is a mixed one with the density matrix
\begin{equation}
    \rho_{\be\be'}=\frac{(2\pi)^3}{V}\frac{(1+\bs\zeta(\spk_2;\spk_2')\bs\s)_{\la_2\la_2'}}{2}\rho(\spk_2;\spk_2').
\end{equation}
Supposing that $\rho(\spk_2,\spk_2')$ is different from zero only in a small vicinity of the diagonal, we can write
\begin{equation}\label{dens_matr_h_appr}
    \rho_{\be\be'}\approx\frac{(2\pi)^3}{V}\frac{(1+\bs\zeta\bs\s)_{\la_2\la_2'}}{2}\rho(\spk_2;\spk_2'),
\end{equation}
where $\bs\zeta:=\bs\zeta(\spk_2;\spk_2)$. We also assume that the detector records plane-wave photons with the momentum $\spk_4$. In this case, the expression standing at the projector $D_{\bar{\be}'\bar{\be}}$ in formula \eqref{PD_fin_mix} for the inclusive probability becomes
\begin{equation}
    \frac{(2\pi)^3}{V}\frac{1}{2} \big\{\rho(1+\bs\zeta\bs\s) +i\vk[(\xi_0+\bs\xi\bs\s)(1+\bs\zeta\bs\s)\tilde{\rho} -c.c.] \big\}_{\la_4\la_4'},
\end{equation}
where $\rho:=\rho(\spk_4;\spk_4)$ and
\begin{equation}
    \tilde\rho=\tilde{\rho}(\spk_{4\parallel};\spk_4)=\int d\spk_{4\perp} \rho(\spk_{4\parallel},\spk_{4\perp};\spk_4')\Big|_{\spk_4'=\spk_4}.
\end{equation}
Let the detector record the hard probe photons in some spin state specified by the projector $D^{(s)}_{\la_4'\la_4}$. Then the inclusive probability \eqref{PD_fin_mix} to record a hard photon in this state is written as
\begin{equation}\label{dP_D_fin}
    dP_D=\frac12 \sum_{\la_4,\la_4'} D^{(s)}_{\la_4'\la_4}  \Big\{\rho -2\vk(\xi_0+\bs\xi\bs\zeta)\im\tilde{\rho} +\big[\rho\bs\zeta -2\vk(\bs\xi+\xi_0\bs\zeta)\im\tilde{\rho} -2\vk(\bs\xi\times\bs\zeta)\re\tilde{\rho} \big]\bs\s \Big\}_{\la_4\la_4'} d\spk_4.
\end{equation}
Recall that this expression is obtained in the leading order of perturbation theory and describes the interference of a free passed wave with its scattered part. This expression is valid only in the parameter domain where the overlap of the interfering waves is substantial. The correction to the trivial (free) contribution to the inclusive probability turns out to be of the order $\al^2$ rather than $\al^4$ as for the standard expression for the light-by-light scattering cross-section \cite{KotSerb97}. Moreover, in deriving expression \eqref{dP_D_fin}, it has been assumed that the wave packet of the probe photon is sufficiently narrow in the momentum space, i.e., $|\De\spk|$ obeys conditions \eqref{condition1} and is much less than the typical scale of variation of the wave function of soft photons $s_\al$. The complex amplitude $s_\al$ describing the state of soft photons has been supposed to be paraxial.

Consider some particular cases of general formula \eqref{dP_D_fin}. If the probe photons are naturally polarized, viz.,  $\bs\zeta=0$, then under the above assumptions formula \eqref{dP_D_fin} implies
\begin{equation}
    dP_D=\frac12 \sum_{\la_4,\la_4'} D^{(s)}_{\la_4'\la_4} \big[\rho -2\vk(\xi_0+\bs\xi\bs\s)\im\tilde{\rho} \big]_{\la_4\la_4'} d\spk_4.
\end{equation}
In this case, the nontrivial contributions to the inclusive probability stem from imaginary part of the density matrix of the hard probe photon in the momentum space. The hard photons being initially in the state \eqref{dens_matr_h_appr} with $\bs\zeta=0$ become polarized with the Stokes vector proportional to the vector $\bs\xi$. In general, the presence of the imaginary contribution to the density matrix of a probe photon gives rise to the following transform of the Stokes parameters:
\begin{equation}\label{de_zeta_im}
    \zeta^0\rightarrow\zeta'^0=\zeta^0 -2\vk(\xi_0+\bs\xi\bs\zeta)\frac{\im\tilde{\rho}}{\rho},\qquad \bs\zeta\rightarrow\bs\zeta'=\bs\zeta -2\vk(\bs\xi+\xi_0\bs\zeta)\frac{\im\tilde{\rho}}{\rho}.
\end{equation}
The imaginary contributions to the density matrix are absent for a usual narrow (in the momentum space) Gaussian wave packet. However, the imaginary part of $\tilde{\rho}$ may appear due to nontrivial structure of the wave packet. For example, such imaginary contributions exist for twisted and Airy states, coherent superposition of several Gaussians, and others (see, e.g., \cite{BliokhVErev,LBThY,SerboNew}). If $\im\tilde{\rho}=0$, then
\begin{equation}
    dP_D=\frac12 \sum_{\la_4,\la_4'} D^{(s)}_{\la_4'\la_4}  \big[\rho +\big(\rho\bs\zeta -2\vk(\bs\xi\times\bs\zeta)\re\tilde{\rho} \big)\bs\s \big]_{\la_4\la_4'} d\spk_4.
\end{equation}
As a result of interaction with photons in the state $s_\al$, the Stokes vector of the probe photon is changed in accordance with the rule (cf. \cite{KotSerb97,Maishev97,Sawyer04,BelMai12,Sawyer15})
\begin{equation}\label{de_zeta}
    \bs\zeta\rightarrow\bs\zeta'=\bs\zeta-2\vk(\bs\xi\times\bs\zeta)\frac{\re\tilde{\rho}}{\rho}.
\end{equation}
As we see, in this case the Stokes vector $\bs\zeta$ precesses around the vector $\bs\xi$. The polarization degree of a hard probe photon, $|\bs\zeta|$, is conserved \cite{KotSerb97,Maishev97,Sawyer04,BelMai12,Sawyer15} up to the terms of higher order in the coupling constant. The precession frequency depends substantially on the form of the density matrix of the probe photon. In the general case described by formula \eqref{dP_D_fin}, the Stokes vector undergoes simultaneous transforms given by \eqref{de_zeta_im} and \eqref{de_zeta}.

Let us estimate a relative magnitude of the quantum corrections in \eqref{dP_D_fin}, \eqref{de_zeta_im}, and \eqref{de_zeta}. By the order of magnitude, the relative value of this correction equals $\eta:=2\vk\xi_0(\s^h_\perp)^2$, where $\s^h_\perp$ is the standard deviation of the transverse momentum component in the wave packet of the probe photon $h_\be$ and $\vk\sim \al^2/2\pi^2|\spk_4|$. Thus,
\begin{equation}
    \eta\sim\frac{\al^2}{\pi^2} f_s(s)\frac{E_s(\s^h_\perp)^2}{\spk_3^2|\spk_4|},
\end{equation}
where $E_s$ is the average energy of photons in the state $s_\al$. For the photons from the state $s_\al$ to participate in the reaction, their wave functions must overlap with the wave function of the probe photon. Therefore, $E_s\sim w_s L/(\s^h_\perp)^2$, where $L$ is the length of the path traveled by the probe photon wave packet in the tested one. As a result,
\begin{equation}
    \eta\sim \frac{\al^2}{\pi^2} f_s(s) \frac{w_sL}{\spk_3^2|\spk_4|}.
\end{equation}
We see from \eqref{chi_est_gen} that by the order of magnitude,
\begin{equation}
    \eta\sim \chi_{ij} |\spk_4|L,
\end{equation}
what is anticipated on physical grounds. Taking the estimate \eqref{chi_est_low} for the susceptibility, we have
\begin{equation}\label{eta_est_gen}
    \eta\sim\frac{\al}{\pi} \frac{\spk_3^2}{m^2} K_u^2 |\spk_4|L=2.31\times10^{-8} K_u^2 \frac{|\spk_4|}{m} \frac{L}{\text{$\mu$m}} \frac{\spk_3^2}{\text{eV}^2}.
\end{equation}
If the energy of a soft photon $|\spk_3|$, the length $L$, and the undulator strength parameter $K_u$ are such that this quantity is of order of unity or larger, then we need to take into account multiple scattering of the hard probe photon on the photons in the state $s_\al$. For large $K_u$, we have to use the Furry picture. As is seen from this estimate, the contribution of the quantum correction can be rather large. For example, putting $L=10$ $\mu$m, $|\spk_4|=2m$, and $|\spk_3|=1$ eV (see, e.g., \cite{Ahmadin22}), we come to
\begin{equation}\label{eta_est_part}
    \eta\sim 4.61\times 10^{-7} K_u^2.
\end{equation}
The change of the probe photon Stokes vector can be measured by the gamma-ray polarimetry \cite{BCKLP22}.

As for the wave packet of a single photon in the state $s_\al$, we have $w_s\sim |\spk_3|n_s\sim|\spk_3|\s_s^3$ and $L\sim1/\s_s$, where $\s_s$ is the standard deviation of momenta in the wave packet $s_\al$. Hence, for $s\sim4m^2$, we deduce
\begin{equation}
    \eta\sim 5.30\times10^{-6}\frac{\s_s^2}{m^2}.
\end{equation}
For $s\ll4 m^2$, we come to
\begin{equation}
    \eta\sim 6.60\times10^{-7} \frac{s\s_s^2}{m^4}.
\end{equation}
As expected, the interference effect caused by scattering of a photon by a photon is very small in this case.

\section{Conclusion}

Let us sum up the results. We have considered the interference effect in photon-photon scattering where the free passed part of the wave function interferes with its scattered part. The forms of the wave packets of the probe photon and of the tested photon have been fully taken into account. We have restricted our considerations to the case where the probe photon is hard and its state is described by some density matrix, whereas the tested photons are soft and are prepared in some one particle or coherent states. Only the leading contributions of the perturbation theory to the inclusive probability to record the probe photon have been retained. In the case we have considered, these contributions stand at zeroth and second powers of the fine structure constant $\al$ \cite{KotSerb97} and, in fact, describe the evolution of the probe photon wave function traversing an effective dispersive medium represented by the soft tested photons. Notice that the standard leading contribution to the cross-section of light-by-light scattering is of order $\al^4$, and we have not taken into account this contribution. Moreover, we have supposed that the total energy of the probe and tested photons is below the electron-positron pair creation threshold, i.e., the abovementioned medium is transparent.

If the wave function of hard probe photon is sufficiently narrow in the momentum space then it is reasonable to employ the small recoil approximation in considering the interference effect. Using this approximation, we have obtained the general and rather compact expression \eqref{ch_fin} for the on-shell susceptibility tensor of a beam of photons and of a single photon wave packet. This tensor describes a birefringent gyrotropic dispersive medium. At the small probe photon momenta, it takes a finite nonzero value and gyrotropy disappears. In increasing the probe photon momenta, the components of the susceptibility tensor rapidly increase and at the electron-positron pair creation threshold gyrotropy becomes of the same order of magnitude as the other contributions to the susceptibility tensor. We have found the estimates for the order of magnitude of the susceptibility tensor in different regimes. Furthermore, using the formalism developed in Sec. \ref{Suscept_Phot}, in Appendix \ref{Sus_Sing_Electr} we have generalized the expression for the susceptibility tensor of a single electron wave packet derived in \cite{KazSol22} to a nonstationary case.

Assuming that the recoil momentum is much less than the typical scale of variation of the tested photon wave functions in the momentum space and that the state of tested photons is paraxial, we have simplified the general expression for the inclusive probability to record a probe photon to formula \eqref{dP_D_fin}. This formula shows that, in passing through the effective medium, the Stokes parameters of the probe photon change and this effect is rather large for a strong beam of tested photons that is sufficiently wide in space. We have found formulas \eqref{de_zeta_im} and \eqref{de_zeta} for the evolution of the Stokes parameters that generalize the analogous expression obtained in \cite{KotSerb97,Maishev97,Sawyer04,BelMai12,Sawyer15} for plane waves. It appears the evolution of the Stokes vector strongly depends on the form of the probe photon wave packet. We have provided the estimates for the order of magnitude of this effect in various regimes. The estimates \eqref{eta_est_gen}, \eqref{eta_est_part} show that this effect can be observed at present and planned facilities provided the control of polarization of hard probe photon and of its wave packet profile is possible \cite{NakHom17,Takahashi18,EMRY22,Karb22,Budker22}.

\paragraph{Acknowledgments.}

We appreciate the anonymous referee for valuable comments. This study was supported by the Tomsk State University Development Programme (Priority-2030).

\appendix
\section{Traces}\label{App_Traces}

In deriving the general expression for the inclusive probability to record a photon, it is necessary to evaluate the traces of operators \eqref{traces}. Let us present here some details of these calculations. As regards the first trace, we obtain
\begin{equation}
    \Sp(\hat{R}_{ph}\hat{\tilde{\Pi}}_{D})=\lan\bar{d}|(1-:\exp(-\hat{c}^\dag D\hat{c}):)|d\ran e^{-\bar{d}d}=1-e^{-\bar{d}Dd},
\end{equation}
where we have used the fact that the coherent state,
\begin{equation}
    |d\ran=e^{d\hat{c}^\dag}|0\ran,
\end{equation}
is an eigenvector for the annihilation operator $\hat{c}_\be$ with the eigenvalue $d_\be$. As far as the second trace is concerned, we have
\begin{equation}
\begin{split}
    \Sp(\hat{R}_{ph}\hat{\tilde{\Pi}}_{D}\hat{C})&=e^{-\bar{d}d}\lan\bar{d}|(1-:\exp(-\hat{c}^\dag D\hat{c}):)\hat{c}^\dag_{\bar{\al}} \hat{c}^\dag_{\bar{\be}} C_{\bar{\al}\bar{\be}\al\be} \hat{c}_{\al}\hat{c}_{\be} |d\ran=\\
    &=e^{-\bar{d}d} d_\al d_\be C_{\bar{\al}\bar{\be}\al\be} \frac{\de}{\de d_{\bar{\al}}} \frac{\de}{\de d_{\bar{\be}}} \lan\bar{d}|(1-:\exp(-\hat{c}^\dag D\hat{c}):) |d\ran=\\
    &=\big[\bar{d}_{\bar{\al}} \bar{d}_{\bar{\be}} -(\bar{d}\tilde{D})_{\tilde{\al}}(\bar{d}\tilde{D})_{\bar{\be}} e^{-\bar{d}Dd} \big]C_{\bar{\al}\bar{\be}\al\be}d_\al d_\be.
\end{split}
\end{equation}

\section{Susceptibility of a single electron wave packet}\label{Sus_Sing_Electr}

In the paper \cite{KazSol22}, the explicit expression for the on-shell susceptibility tensor of a single electron wave packet was obtained. In deriving this expression, certain approximations were made that, in particular, allowed one to consider the electron wave packet as some stationary medium. The last condition can be relaxed by conducting the calculations along lines of Sec. \ref{Suscept_Phot}. In this Appendix, we provide a brief derivation of the expression for the susceptibility tensor of an electron wave packet in a nonstationary case.

In the paper \cite{KazSol22}, formula (106) was derived for the matrix $\Phi_{\bar{\be}\be}$ in the limit of a small recoil $\De\spk$. It reads
\begin{equation}\label{Phi_electr}
    \Phi(\la',\spk';\la,\spk)=-2\pi ie^2\frac{\bar{e}^{(\la')}_i(\spk') e^{(\la)}_i(\spk)}{2V\sqrt{k_0'k_0}} \sum_s\int\frac{d\spp}{E(\spp)} \de(p_0+k_0-p_0'-k_0')\sum_{N=1}^\infty\rho^{(N,1)}_{ss}(\spp,\spp-\De\spk).
\end{equation}
Representing the delta function as
\begin{equation}
    \de(p_0+k_0-p_0'-k_0')=\int\frac{d x^0}{2\pi} e^{-i(p_0+k_0-p_0'-k_0')x^0},
\end{equation}
introducing the density matrix at the instant of time $x^0$,
\begin{equation}
    \rho(\spp,\spp-\De\spk;x^0):=e^{-ik_0x^0}\rho(\spp,\spp-\De\spk)e^{ik_0'x^0},
\end{equation}
and the relativistic density matrix in the coordinate representation,
\begin{equation}
    \rho(\spx,\spy;x^0):=\int\frac{d\spp d\spp' m}{(2\pi)^3\sqrt{E(\spp)E(\spp')}}e^{i\spp\spx-i\spp'\spy} \rho(\spp,\spp';x^0),
\end{equation}
it is not difficult to cast expression \eqref{Phi_electr} into the form
\begin{equation}
    \Phi(\la',\spk';\la,\spk)=- ie^2\frac{\bar{e}^{(\la')}_i(\spk') e^{(\la)}_i(\spk)}{2mV\sqrt{k_0'k_0}} \int d^4x e^{i(k'-k)_\mu x^\mu} \rho(\spx,\spx;x^0),
\end{equation}
where
\begin{equation}
    \rho(\spx,\spx;x^0):=\sum_s\sum_{N=1}^\infty N\rho^{(N,1)}_{ss}(\spx,\spx;x^0).
\end{equation}
Comparing the expression for $\Phi_{\bar{\be}\be}$ with the amplitude of scattering by a dielectric medium \eqref{S_matr_chi}, we conclude that, in the small recoil limit, the susceptibility tensor turns out to be
\begin{equation}\label{chi_elect}
    \chi_{ij}(x;\spK)=-\frac{4\pi\al\rho(\spx,\spx;x^0)}{m K^2_0}\de_{ij}.
\end{equation}
This expression coincides with the susceptibility tensor of an electron plasma. Notice that formula \eqref{chi_elect} is also valid for a single electron wave packet \cite{KazSol22}.


\end{document}